\documentclass[aps,prl,
superscriptaddress,
showpacs,
preprintnumbers,
bibnotes,
amsmath,
amssymb,
twocolumn,
floatfix,
]{revtex4-1}

\usepackage{graphicx}
\usepackage{dcolumn}
\usepackage{bm}
\usepackage[usenames,dvipsnames]{color}
\usepackage{verbatim}
\usepackage{amsfonts}
\usepackage{longtable}
\usepackage{natbib}
\usepackage{hyperref}

\usepackage{dcolumn}

\DeclareMathOperator{\im}{Im}
\DeclareMathOperator{\tr}{Tr}
\newcolumntype{d}[1]{D{;}{.}{#1}}

\graphicspath{{figs/}}

\def\slashed#1{\kern+0.10em /\kern-0.50em #1}

\newcommand{\figscale}{0.53}
\begin{document}
\title{Calculation of the hadronic vacuum polarization disconnected contribution to the muon anomalous magnetic moment}

\newcommand\bnl{Physics Department, Brookhaven National Laboratory, Upton, NY 11973, USA}
\newcommand\cu{Physics Department, Columbia University, New York, NY 10027, USA}
\newcommand\pu{School of Computing \& Mathematics, Plymouth University, Plymouth PL4 8AA, UK}
\newcommand\riken{RIKEN-BNL Research Center, Brookhaven National Laboratory, Upton, NY 11973, USA}
\newcommand\edinb{SUPA, School of Physics, The University of Edinburgh, Edinburgh EH9 3JZ, UK}
\newcommand\uconn{Physics Department, University of Connecticut, Storrs, CT 06269-3046, USA}
\newcommand\soton{School of Physics and Astronomy, University of Southampton,  Southampton SO17 1BJ, UK}
\newcommand\york{Mathematics \& Statistics, York University, Toronto, ON, M3J 1P3, Canada}
\newcommand\cssm{CSSM, University of Adelaide, Adelaide 5005 SA, Australia}
\newcommand\cern{CERN, Physics Department, 1211 Geneva 23, Switzerland}

\author{T.~Blum}\affiliation{\uconn}
\author{P.A.~Boyle}\affiliation{\edinb}
\author{T.~Izubuchi}\affiliation{\bnl}\affiliation{\riken}
\author{L.~Jin}\affiliation{\cu}
\author{A.~J\"uttner}\affiliation{\soton}
\author{C.~Lehner}\thanks{Corresponding author}\email{clehner@quark.phy.bnl.gov}\affiliation{\bnl}
\author{K.~Maltman}\affiliation{\york}\affiliation{\cssm}
\author{M.~Marinkovic}\affiliation{\cern}
\author{A.~Portelli}\affiliation{\edinb}\affiliation{\soton}
\author{M.~Spraggs}\affiliation{\soton}

\collaboration{RBC and UKQCD Collaborations}
\noaffiliation

\date{December 30, 2015}

\begin{abstract} 
  We report the first lattice QCD calculation of the hadronic vacuum
  polarization disconnected contribution to the muon anomalous
  magnetic moment at physical pion mass.  The calculation uses a
  refined noise-reduction technique which enabled the control of
  statistical uncertainties at the desired level with modest
  computational effort.  Measurements were performed on the $48^3
  \times 96$ physical-pion-mass lattice generated by the RBC and UKQCD
  collaborations.  We find $a_\mu^{\rm HVP~(LO)~DISC} = -9.6(3.3)(2.3)
  \times 10^{-10}$, where the first error is statistical and the
  second systematic.
\end{abstract}


\preprint{}

\maketitle

\section{Introduction}
The anomalous magnetic moment of leptons provides a powerful tool to
test relativistic quantum-mechanical effects at tremendous precision.
Consider the magnetic dipole moment of a fermion
\begin{align}
  \vec{\mu} = g
\left( \frac{e}{2m} \right) \vec{s} \,,
\end{align}
where $\vec{s}$ is the particle's spin, $e$ is its charge, and $m$ is
its mass.  While Dirac's relativistic quantum-mechanical treatment of
a fermion coupled minimally to a classical photon background predicts
a Land\'e factor of $g=2$, additional electromagnetic quantum effects
allow the anomalous magnetic moment $a=(g-2)/2$ to assume a non-zero
value.  These anomalous moments are measured very precisely.  For the
electron, e.g., $ a_e = 0.00115965218073(28)$ \cite{Hanneke:2008tm}
yielding the currently most precise determination of the fine
structure constant $ \alpha= 1/137.035999157(33)$ based on a 5-loop
quantum electrodynamics (QED) computation \cite{Aoyama:2014sxa}.

The muon anomalous magnetic moment promises high sensitivity to new
physics (NP) beyond the standard model (SM) of particle physics.  In
general, new physics contributions to $a_\ell$ are expected to scale
as $ a_\ell - a_\ell^{\rm SM} \propto (m^2_\ell / \Lambda^2_{\rm NP})$
for lepton $\ell=e,\mu,\tau$ and new physics scale $\Lambda_{\rm NP}$.
With $a_\tau$ being currently experimentally inaccessible, $a_\mu$ is
the optimum channel to uncover new physics.

Interestingly, there is an $3.1$--$3.5 \sigma$ tension between
current experimental and theoretical determinations of $a_\mu$,
\begin{align}
  a_\mu^{\rm EXP} - a_\mu^{\rm SM} = \,\,& (27.6 \pm 8.0) \times 10 ^ {-10}~\text{\cite{Davier:2010nc}} \,, \notag\\
  &(25.0 \pm 8.0) \times 10 ^ {-10}~\text{\cite{Hagiwara:2011af}} \,,
\end{align}
where the experimental measurement is dominated by the BNL experiment
E821 \cite{Bennett:2006fi}.  The theoretical prediction
\cite{PDG2013} is broken down in individual contributions in
Tab.~\ref{tab:mugmtwo}.

The theory error is dominated by the hadronic
vacuum polarization (HVP) and hadronic light-by-light (HLbL)
contributions.  A careful first-principles determination of these
hadronic contributions is very much desired to resolve or more firmly
establish the tension with the current SM prediction.  Furthermore,
the future $a_\mu$ experiments at Fermilab (E989) \cite{Carey:2009zzb}
and J-PARC (E34) \cite{Aoki:2009xxx} intend to decrease the
experimental error by a factor of four.  Therefore a similar reduction
of the theory error is essential in order to make full use of the
experimental efforts.

\begin{table}[tb]
  \centering
  \begin{tabular}{lrr}\hline\hline
    Contribution & Value $\times 10^{10}$ & Uncertainty $\times 10^{10}$\\\hline
    QED & 11 658 471.895 & 0.008 \\
    Electroweak Corrections & 15.4 & 0.1 \\
    HVP (LO) \cite{Davier:2010nc} & 692.3  & 4.2 \\
    HVP (LO) \cite{Hagiwara:2011af} & 694.9  & 4.3 \\
    HVP (NLO) & -9.84 & 0.06 \\
    HVP (NNLO) & 1.24 & 0.01 \\
    HLbL & 10.5 & 2.6 \\
    \hline
    Total SM prediction \cite{Davier:2010nc} & 11 659 181.5 & 4.9 \\
    Total SM prediction \cite{Hagiwara:2011af} & 11 659 184.1 & 5.0 \\    \hline
    BNL E821 result & 11 659 209.1 & 6.3 \\
    Fermilab E989 target & & $\approx$  1.6\\\hline\hline
  \end{tabular}
  \caption{Current standard model prediction of $a_\mu$ including uncertainties
    contrasted with the experimental target precision of the upcoming Fermilab E989 experiment \cite{PDG2013,Grange:2015fou,Kurz:2014wya}. 
    The individual contributions are defined in Ref.~\cite{PDG2013}.}
  \label{tab:mugmtwo}
\end{table}

The current SM prediction for the HLbL \cite{Prades:2009tw} is based
on a model of quantum chromodynamics (QCD), however, important
progress towards a first-principles computation has been made recently
\cite{Blum:2015gfa,Green:2015sra,Blum:2014oka}.  The uncertainty of the HVP
contribution may be reduced to $\delta a_\mu = 2.6
\times 10^{-10}$ using improved experimental $e^+e^-$ scattering data
\cite{Grange:2015fou}.  An ab-initio theory prediction based on QCD,
however, can provide an important alternative determination that is
systematically improvable to higher precision.

One of the main challenges in the first-principles computation of the
HVP contribution with percent or sub-percent uncertainties is the
control of statistical noise for the quark-disconnected contribution
(see Fig.~\ref{fig:classificationhvp}) at physical pion mass.
Significant progress has been made recently in the computation of an
upper bound \cite{Francis:2014hoa,Bali:2015msa,Burger:2015oya}, an
estimate using lattice QCD data at heavy pion mass
\cite{Chakraborty:2015ugp}, and towards a first-principles computation
at physical pion mass \cite{BMWDiscoLattice2015}.  Here we present the
first result for $a_\mu^{\rm HVP~(LO)~DISC}$ at physical pion mass.
We report the result for the combined up, down, and strange quark
contributions.

\section{Computational Method}
In the following we describe the refined noise-reduction
technique that allowed for the control of the statistical
noise with modest computational effort.

\begin{figure}[tb]
  \centering
  \includegraphics[height=3.5cm]{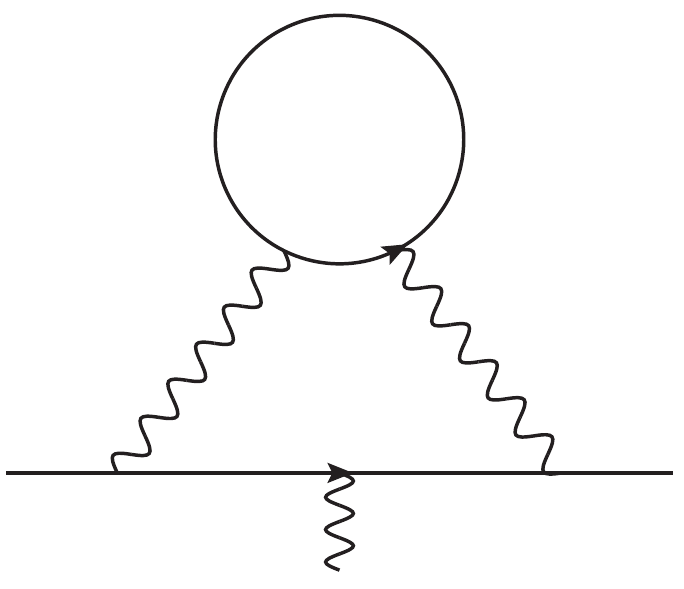} \hspace{0.3cm}
  \includegraphics[height=3.5cm]{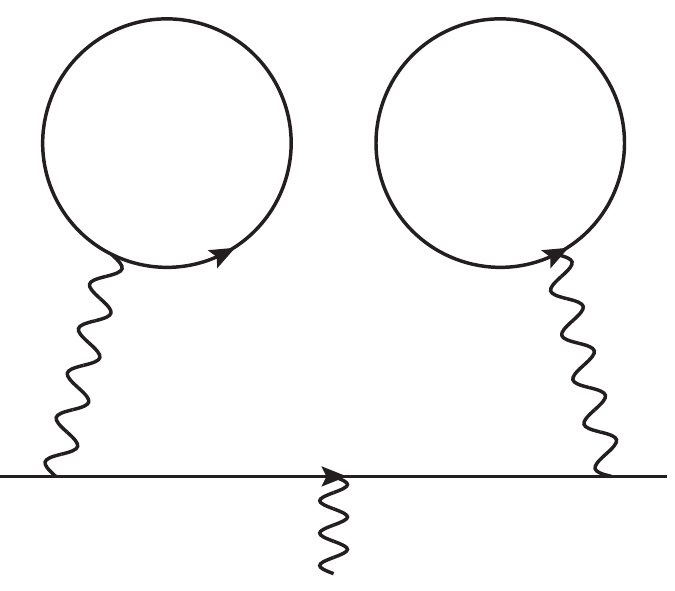} \hspace{0.3cm}
  \caption{HVP contributions to $a_\mu$ with external photon attached
    to the muon line.  As common for non-perturbative lattice QCD
    computations, one does not explicitly draw gluons but understands
    each diagram to stand for all orders in QCD.  
}
  \label{fig:classificationhvp}
\end{figure}

We follow the basic steps of Ref.~\cite{Blum:2002ii} and treat the muon
and photon parts of the diagrams in Fig.~\ref{fig:classificationhvp}
analytically, writing
\begin{align}\label{eqn:defamu}
  a_\mu = 4 \alpha^2 \int_0^\infty d(q^2) f(q^2) [ \Pi(q^2) - \Pi(q^2=0) ] \,,
\end{align}
where $f(q^2)$ is a known analytic function \cite{Blum:2002ii} and
$\Pi(q^2)$ is defined in the continuum through the two-point function
\begin{align}
  \sum_x e^{i q x} \langle J_\mu(x) J_\nu(0) \rangle = (\delta_{\mu\nu} q^2 - q_\mu q_\nu) \Pi(q^2)
\end{align}
with sum over space-time coordinate $x$ and
$ J_\mu(x) = i\sum_f Q_f \overline{\Psi}_f(x) \gamma_\mu \Psi_f(x)$.
The sum is over quark flavors $f$ with QED charge $Q_f$ ($Q_u=2/3$, $Q_{d/s}=-1/3$).
We compute $\Pi(q^2)$ using the kernel
function of Refs.~\cite{Bernecker:2011gh,Feng:2013xsa}
\begin{align}\label{eqn:wick}
  \Pi(q^2) - \Pi(q^2=0) &= \sum_{t} \left( \frac{\cos(q t) - 1}{q^2} + \frac12 t^2\right) C_{\rm all}(t)
\end{align}
with
\begin{align}\label{eqn:ct}
  C_{\rm all}(t) = \frac13 \sum_{\vec{x}}\sum_{j=0,1,2} \langle J_j(\vec{x},t) J_j(0) \rangle
\end{align}
which sufficiently suppresses the short-distance contributions such
that we are able to use two less computationally costly,
non-conserved, local lattice vector currents \footnote{The appropriate
  normalization factors of $Z_V^2$ are of course included in our
  computation.}.  For convenience, we have split the space-time sum
over $x$ in a spatial sum over $\vec{x}$ and a sum over the time
coordinate $t$.  We sum over spatial Lorentz indices $0,1,2$.

The Wick contraction in Eq.~\eqref{eqn:ct} yields both connected and
disconnected diagrams of Fig.~\ref{fig:classificationhvp}.  In the
following $C$ stands for the combined up-, down-, and strange-quark
disconnected contribution of $C_{\rm all}$, while $C_s$ stands for
the strange-quark connected contribution of $C_{\rm all}$.  The reason
for defining $C_s$ will become apparent below.  The light up and
down flavors are treated as mass degenerate such that
\begin{align}\label{eqn:concrete}
C(t) =\frac1{3V}\sum_{j=0,1,2} \sum_{t'} \langle {\cal V}_j(t + t') {\cal V}_j(t') \rangle_{\rm SU(3)}
\end{align}
where $V$ stands for the four-dimensional lattice volume, ${\cal
  V}_\mu = (1/3) ( {\cal V}_\mu^{u/d} - {\cal V}_\mu^s )$, the average
is over all SU(3) gauge configurations, and
\begin{align}
  {\cal V}^f_\mu(t) = \sum_{\vec{x}} \im\tr[ D^{-1}_{\vec{x},t ; \vec{x}, t}(m_f) \gamma_\mu ]
\end{align}
with Dirac operator $D(m_f)$ evaluated at quark mass $m_f$.

\begin{figure}[tb]
\begin{center}
\includegraphics[scale=\figscale,page=1]{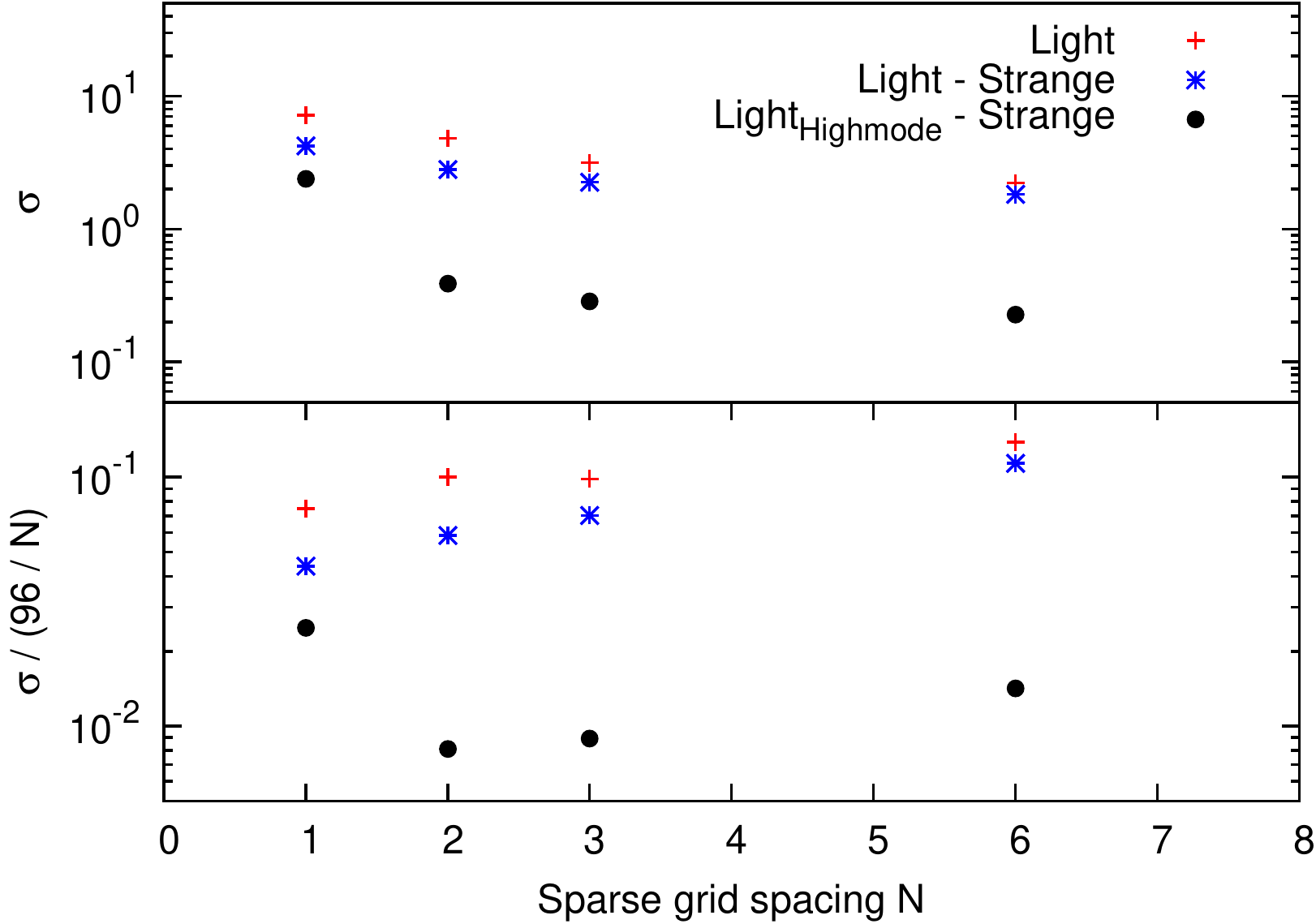}
\end{center} 
\caption{Noise of single vector operator loop as a function of sparse
  grid spacing $N$.  The figure at the bottom normalizes the noise by
  taking into account the additional volume averaging for smaller
  values of $N$.}
\label{fig:noise}
\end{figure}

Controlling statistical fluctuations is the largest challenge in the
computation of the disconnected contribution.  In order to
successfully measure the disconnected contribution, two
conditions need to be satisfied: (i) large fluctuations of the up/down
and strange contributions that enter with opposite sign need to cancel
\cite{Francis:2014hoa} and (ii) the measurement needs to average over
the entire spacetime volume without introducing additional noise.  Here
we use the following method to satisfy both (i) and (ii)
simultaneously.  First, the full quark propagator is separated in high
and low-mode contributions, where the former are estimated
stochastically and the latter are averaged explicitly
\cite{Foley:2005ac}, i.e., we separate
$ D^{-1} = \sum_n v^n (w^n)^\dagger + D^{-1}_{\rm high}$,
where the vectors $v^n$ and $w^n$ are reconstructed from the even-odd
preconditioned low-modes $n$ of the Dirac propagator $D^{-1}$.  It is
now crucial to include all modes with eigenvalues up to the strange
quark mass in the set of low modes for the up and down quark
propagators to satisfy (i).  Since the signal is the difference of
light and strange contributions, we may then expect the high-mode
contribution to be significantly suppressed and the low-mode
contribution to contain the dominant part of the signal.  This is
indeed the case in our computation and yields a substantial
statistical benefit since we evaluate the low modes exactly without
the introduction of noise and average explicitly over the entire
volume.

In order to satisfy (ii), we must control the stochastic noise of the
high-mode contributions originating from unwanted long-distance
contributions of the random $Z_2$ sources of Ref.~\cite{Foley:2005ac}.
We achieve this by using what we refer to as sparsened $Z_2$ noise
sources which have support only for points $x_\mu$ with $(x_\mu -
x_\mu^{(0)}) \mbox{ mod } N = 0$ thereby defining a sparse grid with
spacing $N$, similar to Ref.~\cite{Li:2010pw}.  While a
straightforward dilution strategy \cite{Foley:2005ac} would require us
to sum over all possible offsets of the sparse noise grid,
$x_\mu^{(0)}$, we choose the offset stochastically for each individual
source which allows us to project to all momenta.  It also allows us
to avoid the largest contribution of such random sources to the noise
which comes from random sources at nearby points.

The parameter choice of $N$ is crucial to satisfy (ii) with minimal
cost.  Figure~\ref{fig:noise} shows the square root of the variance
$\sigma^2$ of ${\cal V}$ on a single lattice configuration over time
coordinate $t$ and Lorentz index $\mu$.  Since we can use all possible
O$(M^2)$ combinations of $M$ high-mode sources and time-coordinates in
Eq.~\eqref{eqn:concrete}, we may expect a noise suppression of
O$(1/M)$ as long as individual contributions are sufficiently
statistically independent.  A similar idea of O$(1/M)$ noise reduction
was recently successfully used in Ref.~\cite{Blum:2015gfa}.  We find
this to hold to a large degree, and therefore also show the
appropriately rescaled $\sigma$ in the lower panel of
Fig.~\ref{fig:noise}.  The figure illustrates the powerful
cancellation of noise between the light and strange quark
contributions and the success of the sparsening strategy.  We find an
optimum value of $N=3$ for the case at hand, which is used for the
subsequent numerical discussion.

We use $45$ stochastic high-modes per configuration and measure on
$21$ Moebius domain wall \cite{Brower:2004xi} configurations of the $48^3 \times 96$
ensemble at physical pion mass and lattice cutoff $a^{-1} = 1.73$ GeV
generated by the RBC and UKQCD collaborations \cite{Blum:2014tka}.
For this number of high modes we find the QCD gauge noise to dominate
the uncertainty for $a^{\rm HVP~(LO)~DISC}_\mu$.  The AMA strategy
\cite{Blum:2012uh,Shintani:2014vja} was employed to reduce the cost of
computing multiple sources on the same configuration.  The computation
presented in this manuscript uses 2000 zMobius
\cite{ChulwooLattice2015} eigenvectors generated as part of an
on-going HLbL lattice computation \cite{Blum:2015gfa}.  We treat the
shorter directions with 48 points as the time direction and average
over the three symmetric combinations to further reduce stochastic
noise.

\section{Analysis and results}
Combining Eqs.~\eqref{eqn:defamu} and \eqref{eqn:wick} and using $C(t) = C(-t)$,
\begin{align}\label{eqn:am}
  a^{\rm HVP~(LO)~DISC}_\mu &= \sum_{t=0}^\infty w_t C(t) 
\end{align}
with appropriately defined $w_t$.  Due to our choice of relatively
short time direction with 48 points, special care needs to be taken to
control potentially missing long-time contributions in $C(t)$.
In the following we estimate these effects quantitatively.  Consider
the vector operator
\begin{align}
  V^{f,f'}_\mu(x) = \overline{\Psi}_f(x) \gamma_\mu \Psi_{f'}(x)
\end{align}
with $f$ and $f'$ denoting quark flavors.  Then
the Wick contractions
\begin{align}
  \langle V^{u,u}_\mu V^{u,u}_\nu \rangle
  -\langle V^{u,d}_\mu V^{d,u}_\nu \rangle
\end{align}
isolate the light-quark disconnected contribution in the isospin
symmetric limit, see also Ref.~\cite{DellaMorte:2010aq}.
Unfortunately there is no similar linear combination (without partial
quenching) that allows for the isolation of the strange-quark
disconnected contribution.  Nevertheless, using
\begin{align}\label{eqn:ww}
  \langle (V^{u,u}_\mu - V^{s,s}_\mu)(V^{u,u}_\nu - V^{s,s}_\nu) \rangle - 
  \langle V^{u,d}_\mu V^{d,u}_\nu \rangle
\end{align}
one can isolate the sum of $C(t)+C_s(t)$, again making use of the
isospin symmetry.  Since this sum corresponds to a complete set of
Feynman diagrams resulting from the above Wick contractions, we can
represent it as a sum over individual exponentials $C(t)+C_s(t) =
\sum_m c_m e^{-E_m t}$ with $c_m \in \mathbb{R}$ and $E_m \in
\mathbb{R}^+$.  The coefficients $c_m$ can be negative because
positivity arguments only apply to some individual Wick contractions in
Eq.~\eqref{eqn:ww} but not necessarily to the sum.

\begin{figure}[tb]
\begin{center}
\includegraphics[scale=\figscale,page=5]{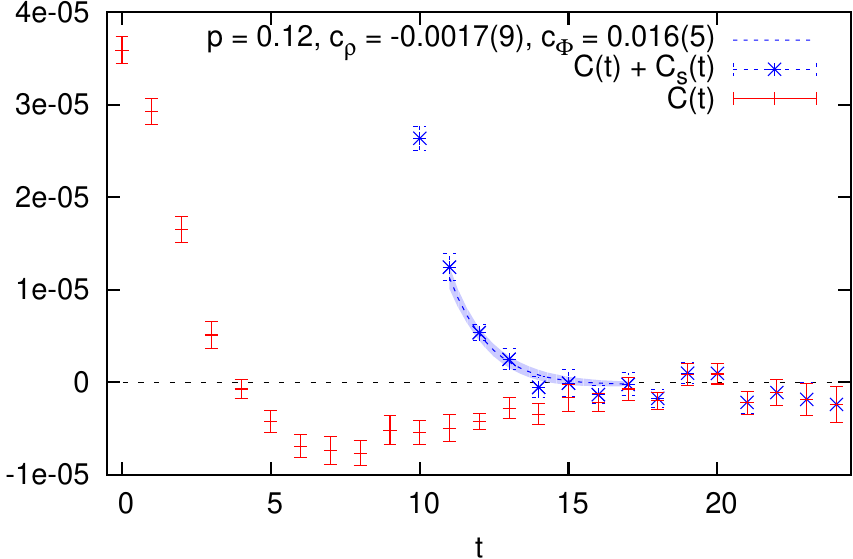}
\end{center} 
\caption{Zero-momentum projected correlator $C(t)$ and $C(t)+C_s(t)$.
  A correlated fit of $\rho(770)$ and $\phi(1020)$ exponentials via $c_\rho
  e^{-E_\rho t} + c_\phi e^{-E_\phi t}$ in the region
  $t\in[11,\ldots,17]$ to $C(t) + C_s(t)$ yields a $p$-value of 0.12.
  We use fixed energies $E_\rho=770$ MeV and $E_\phi=1020$ MeV and fit
  parameters $c_\rho$ and $c_\phi$.}
\label{fig:cor}
\end{figure}

\begin{figure}[tb]
\begin{center}
\includegraphics[scale=\figscale,page=6]{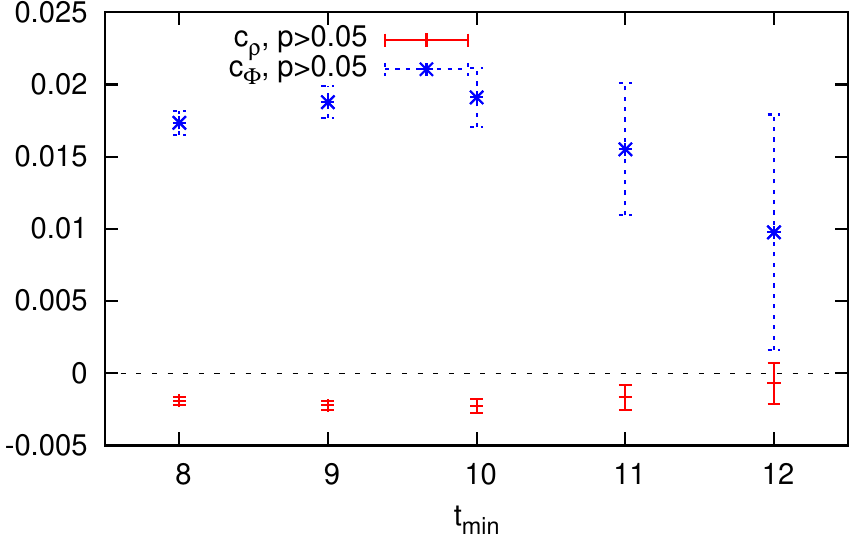}
\end{center} 
\caption{Coefficients and $p$-values of a fit of $c_\rho e^{-E_\rho t}
  + c_\phi e^{-E_\phi t}$ to $C(t)+C_s(t)$ in the region $t\in[t_{\rm
    min},\ldots,17]$.}
\label{fig:fitrangestudy}
\end{figure}

We show $C(t)$ and $C_s(t)$ obtained in our lattice QCD computation in
Fig.~\ref{fig:cor}.  Starting from time-slices 17, 18 the correlator
$C(t)$ is not well resolved from zero, however, from time-slices 11 to
17 a two-state fit including the $\rho(770)$ and $\phi(1020)$
describes $C(t) + C_s(t)$ well.  Here the $\rho$ is a proxy for
combined $\rho$ and $\omega$ contributions due to their similar
energy.  Since these states are not stable in our lattice simulation,
however, this representation using individual exponentials only serves
as a model that fits the data well.  Since this model will only enter
our systematic error estimate, we find this imperfection to be
acceptable.  A systematic study of different fit ranges is presented
in Fig.~\ref{fig:fitrangestudy}, where p-values greater than $0.05$
are found for all fit-ranges $t\in [t_{\rm min},\ldots,17]$ with
$t_{\rm min}\in [8,\ldots,12]$.

We now define the partial sums
\begin{align}\label{eqn:lt}
  L_T &= \sum_{t=0}^T w_t C(t)\,, \\ \label{eqn:ft}
  F_T(r) &= \sum_{t=T+1}^{t_{\rm max}} w_t (c^r_\rho e^{-E_\rho t}
  + c^r_\phi e^{-E_\phi t} - C_s(t)) \,,
\end{align}
where $c^r_\rho$ and $c^r_\phi$ are the parameters of the fit with
fit-range $r$ and $t_{\rm max}=24$ for our setup.  For sufficiently
large $T$, $L_T$ is expected to exhibit a plateau region as function
of $T$ from which we can determine $a^{\rm HVP~(LO)~DISC}_\mu$.  The
sum $L_T + F_T$ is also expected to exhibit such a plateau to the
extent that the model in $F_T$ describes the data well.

\begin{figure}[tb]
\begin{center}
\includegraphics[scale=\figscale,page=7]{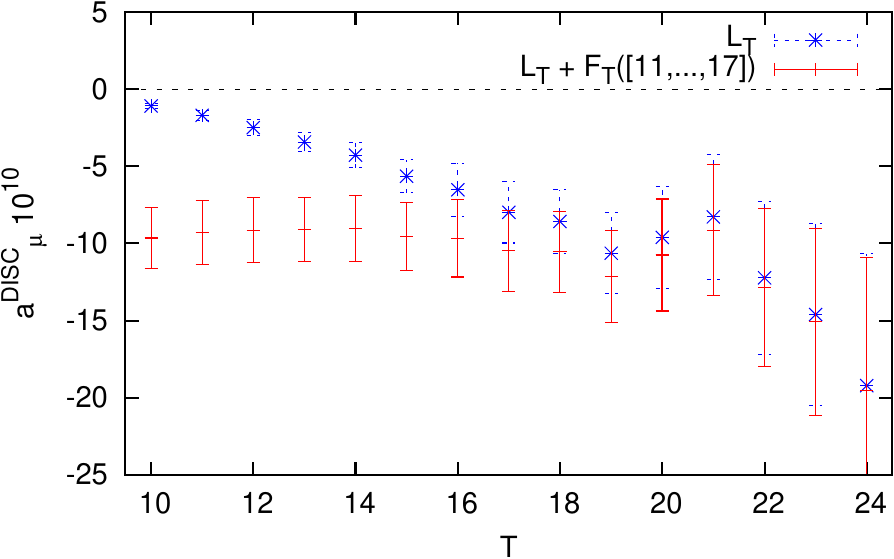}
\vspace{0.1cm}

\includegraphics[scale=\figscale,page=8]{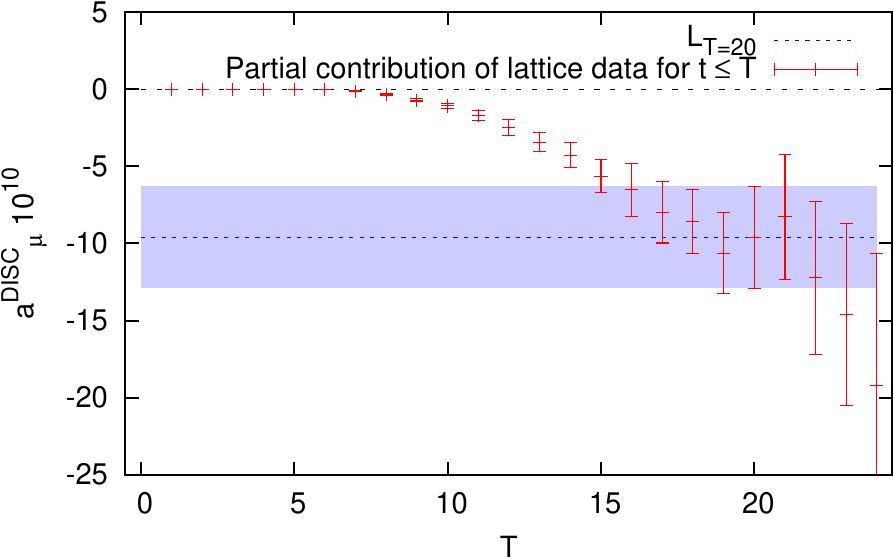}
\end{center} 
\caption{The sum of $L_T$ and $F_T$ defined in Eqs.~\eqref{eqn:lt} and \eqref{eqn:ft} has
  a plateau from which we read off $a^{\rm HVP~(LO)~DISC}_\mu$.  The lower panel compares
  the partial sums $L_T$ for all values of $T$ with our final result
  for $a^{\rm HVP~(LO)~DISC}_\mu$ with its statistical error band.}
\label{fig:result}
\end{figure}

Based on Fig.~\ref{fig:fitrangestudy}, we choose $r=[11,\ldots,17]$ as
preferred fit-range to determine $F_T$ but a cross-check with
$r=[12,\ldots,17]$ has been performed yielding a consistent result.
Figure~\ref{fig:result} shows the resulting plateau-region for $L_T$
and $L_T + F_T$.  In order to avoid contamination of our
first-principles computation with the model-dependence of $F_T$, we
determine $a^{\rm HVP~(LO)~DISC}_\mu$ from $L_{T=20}$ and include
$F_{T=20}$ as systematic uncertainty estimating a potentially missing
long-time tail.  We choose the value at $T=20$ since it appears to be
safely within a plateau region but sufficiently far from $T=24$ to
suppress backwards-propagating effects \footnote{Alternatively taking
  $T=21$ instead of $T=20$ and repeating our procedure to estimate
  systematic uncertainties, we find $a_\mu^{\rm HVP~(LO)~DISC} =
  -8.3(4.0)(1.8)\times 10^{-10}$, where the first error is statistical
  and the second systematic.  This value is consistent with our
  preferred value, however, has a different balance of statistical
  and systematic errors.}.  We find $a^{\rm
  HVP~(LO)~DISC}_\mu=-9.6(3.3)\times 10^{-10}$.

\begin{figure}[tb]
\begin{center}
\includegraphics[scale=\figscale,page=1]{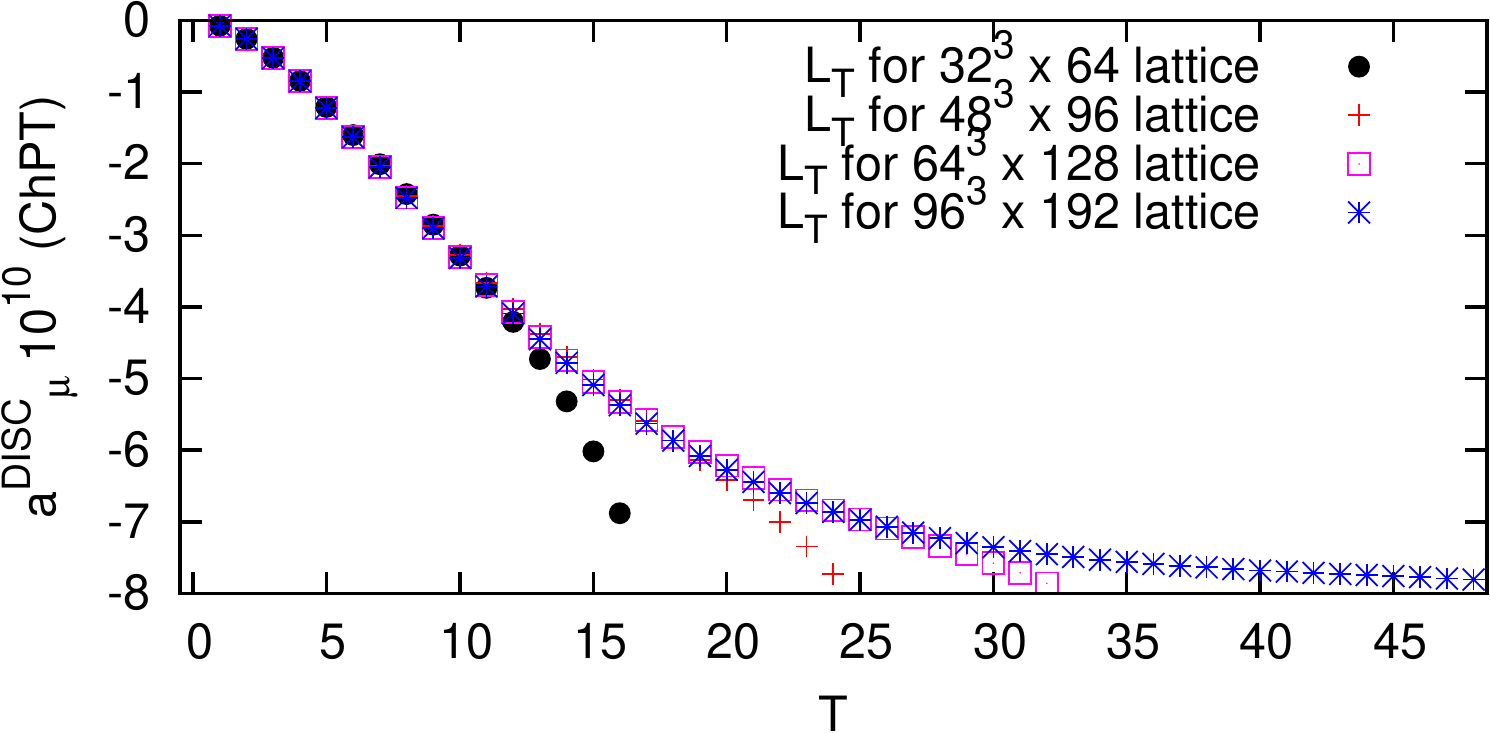}
\end{center} 
\caption{The leading-order pion-loop contribution in finite-volume
  ChPT as function of volume.}
\label{fig:fv}
\end{figure}

We expect the finite lattice spacing and finite simulation volume as
well as long-time contributions to Eq.~\eqref{eqn:am} to dominate the
systematic uncertainties of our result.  With respect to the finite
lattice spacing a reasonable proxy for the current computation may be
our HVP connected strange-quark analysis \cite{MattLattice2015} for
which the $48^3$ result at $a^{-1}=1.73$ GeV agrees within O($5\%$)
with the continuum-extrapolated value.  This is also consistent with a
na\"ive O($a^2\Lambda^2_{\rm QCD}$) power counting, appropriate for
the domain-wall fermion action used here.  The combined effect of the
finite spatial volume and potentially missing two-pion tail is
estimated using a one-loop finite-volume lattice-regulated chiral
perturbation theory (ChPT) version of Eq.~(5.1) of
Ref.~\cite{DellaMorte:2010aq}.  Our ChPT computation also agrees with
Eq.~(2.12) of Ref.~\cite{Aubin:2015rzx} after correcting for a missing
factor of two in the first version of Ref.~\cite{Aubin:2015rzx}.  The
ChPT result is then transformed to position space to obtain $C(t)$.
Fig.~\ref{fig:fv} shows a corresponding study of $L_T$ for different
volumes.  We take the difference of $L_{T=20}$ on the $48^3 \times 96$
lattice used here and $L_{T=48}$ on the $96^3 \times 192$ lattice and
obtain $\delta a^{\rm FV,\pi\pi}_\mu = 1.4 \times 10^{-10}$.  The
remaining long-time effects are estimated by $F_{T=20}$.  We compare
the result for two fit-ranges $F_{T=20}([11,\ldots,17])=-1.1(6) \times
10^{-10}$ and $F_{T=20}([12,\ldots,17])=-0.6(0.9) \times 10^{-10}$.
We conservatively take the one-sigma bound $\delta a^{\rm F_T}=1.7
\times 10^{-10}$ as additional uncertainty.  

Combining the systematic uncertainties in quadrature, we report our
final result
\begin{align}
 a_\mu^{\rm HVP~(LO)~DISC} = -9.6(3.3)(2.3)\times 10^{-10} \,,
\end{align}
where the first error is statistical and the second systematic.

Before concluding, we note that our result appears to be dominated by
very low energy scales.  This is not surprising since the signal is
expressed explicitly as difference of light-quark and
strange-quark Dirac propagators.  We therefore expect energy scales
significantly above the strange mass to be suppressed.  We already
observed this above in the dominance of low modes of the Dirac
operator for our signal.  Furthermore, our result is statistically
consistent with the one-loop ChPT two-pion contribution of
Fig.~\ref{fig:fv}.

\section{Conclusion}
We have presented the first ab-initio calculation of the hadronic
vacuum polarization disconnected contribution to the muon anomalous
magnetic moment at physical pion mass.  We were able to obtain our
result with modest computational effort utilizing a refined
noise-reduction technique explained above.  This computation addresses
one of the major challenges for a first-principles lattice QCD
computation of $a_\mu^{\rm HVP}$ at percent or sub-percent precision,
necessary to match the anticipated reduction in experimental
uncertainty.  The uncertainty of the result presented here is already
slightly below the current experimental precision and can be reduced further
by a straightforward numerical effort.

\section{Acknowledgments}
We would like to thank our RBC and UKQCD collaborators for helpful
discussions and support.  C.L.~is in particular indebted to Norman
Christ, Masashi Hayakawa, and Chulwoo Jung for helpful comments
regarding this manuscript.  This calculation was carried out at the
Fermilab cluster pi0 as part of the USQCD Collaboration.  The
eigenvectors were generated under the ALCC Program of the US DOE on
the IBM Blue Gene/Q (BG/Q) Mira machine at the Argonne Leadership
Class Facility, a DOE Office of Science Facility supported under
Contract De-AC02-06CH11357.  T.B.~is supported by US DOE grant
DE-FG02-92ER40716.  P.A.B.~and A.P.~are supported in part by UK STFC
Grants No. ST/M006530/1, ST/L000458/1, ST/K005790/1, and ST/K005804/1
and A.P.~additionally by ST/L000296/1.  T.I.~and C.L.~are supported in
part by US DOE Contract \#AC-02-98CH10886(BNL).  T.I.~is supported in
part by the Japanese Ministry of Education Grant-in-Aid, No. 26400261.
L.J.~is supported in part by US DOE grant \#de-sc0011941.  A.J.~is
supported by EU FP7/2007-2013 ERC grant 279757.  K.M.~is supported by
the National Sciences and Engineering Research Council of Canada.
M.S.~is supported by EPSRC Doctoral Training Centre Grant
EP/G03690X/1.

\bibliography{refs}

\begin{thebibliography}{34}%
\makeatletter
\providecommand \@ifxundefined [1]{%
 \@ifx{#1\undefined}
}%
\providecommand \@ifnum [1]{%
 \ifnum #1\expandafter \@firstoftwo
 \else \expandafter \@secondoftwo
 \fi
}%
\providecommand \@ifx [1]{%
 \ifx #1\expandafter \@firstoftwo
 \else \expandafter \@secondoftwo
 \fi
}%
\providecommand \natexlab [1]{#1}%
\providecommand \enquote  [1]{``#1''}%
\providecommand \bibnamefont  [1]{#1}%
\providecommand \bibfnamefont [1]{#1}%
\providecommand \citenamefont [1]{#1}%
\providecommand \href@noop [0]{\@secondoftwo}%
\providecommand \href [0]{\begingroup \@sanitize@url \@href}%
\providecommand \@href[1]{\@@startlink{#1}\@@href}%
\providecommand \@@href[1]{\endgroup#1\@@endlink}%
\providecommand \@sanitize@url [0]{\catcode `\\12\catcode `\$12\catcode
  `\&12\catcode `\#12\catcode `\^12\catcode `\_12\catcode `\%12\relax}%
\providecommand \@@startlink[1]{}%
\providecommand \@@endlink[0]{}%
\providecommand \url  [0]{\begingroup\@sanitize@url \@url }%
\providecommand \@url [1]{\endgroup\@href {#1}{\urlprefix }}%
\providecommand \urlprefix  [0]{URL }%
\providecommand \Eprint [0]{\href }%
\providecommand \doibase [0]{http://dx.doi.org/}%
\providecommand \selectlanguage [0]{\@gobble}%
\providecommand \bibinfo  [0]{\@secondoftwo}%
\providecommand \bibfield  [0]{\@secondoftwo}%
\providecommand \translation [1]{[#1]}%
\providecommand \BibitemOpen [0]{}%
\providecommand \bibitemStop [0]{}%
\providecommand \bibitemNoStop [0]{.\EOS\space}%
\providecommand \EOS [0]{\spacefactor3000\relax}%
\providecommand \BibitemShut  [1]{\csname bibitem#1\endcsname}%
\let\auto@bib@innerbib\@empty
\bibitem [{\citenamefont {Hanneke}\ \emph {et~al.}(2008)\citenamefont
  {Hanneke}, \citenamefont {Fogwell},\ and\ \citenamefont
  {Gabrielse}}]{Hanneke:2008tm}%
  \BibitemOpen
  \bibfield  {author} {\bibinfo {author} {\bibfnamefont {D.}~\bibnamefont
  {Hanneke}}, \bibinfo {author} {\bibfnamefont {S.}~\bibnamefont {Fogwell}}, \
  and\ \bibinfo {author} {\bibfnamefont {G.}~\bibnamefont {Gabrielse}},\ }\href
  {\doibase 10.1103/PhysRevLett.100.120801} {\bibfield  {journal} {\bibinfo
  {journal} {Phys.Rev.Lett.}\ }\textbf {\bibinfo {volume} {100}},\ \bibinfo
  {pages} {120801} (\bibinfo {year} {2008})},\ \Eprint
  {http://arxiv.org/abs/0801.1134} {arXiv:0801.1134 [physics.atom-ph]}
  \BibitemShut {NoStop}%
\bibitem [{\citenamefont {Aoyama}\ \emph {et~al.}(2015)\citenamefont {Aoyama},
  \citenamefont {Hayakawa}, \citenamefont {Kinoshita},\ and\ \citenamefont
  {Nio}}]{Aoyama:2014sxa}%
  \BibitemOpen
  \bibfield  {author} {\bibinfo {author} {\bibfnamefont {T.}~\bibnamefont
  {Aoyama}}, \bibinfo {author} {\bibfnamefont {M.}~\bibnamefont {Hayakawa}},
  \bibinfo {author} {\bibfnamefont {T.}~\bibnamefont {Kinoshita}}, \ and\
  \bibinfo {author} {\bibfnamefont {M.}~\bibnamefont {Nio}},\ }\href {\doibase
  10.1103/PhysRevD.91.033006} {\bibfield  {journal} {\bibinfo  {journal} {Phys.
  Rev.}\ }\textbf {\bibinfo {volume} {D91}},\ \bibinfo {pages} {033006}
  (\bibinfo {year} {2015})},\ \Eprint {http://arxiv.org/abs/1412.8284}
  {arXiv:1412.8284 [hep-ph]} \BibitemShut {NoStop}%
\bibitem [{\citenamefont {Davier}\ \emph {et~al.}(2011)\citenamefont {Davier},
  \citenamefont {Hoecker}, \citenamefont {Malaescu},\ and\ \citenamefont
  {Zhang}}]{Davier:2010nc}%
  \BibitemOpen
  \bibfield  {author} {\bibinfo {author} {\bibfnamefont {M.}~\bibnamefont
  {Davier}}, \bibinfo {author} {\bibfnamefont {A.}~\bibnamefont {Hoecker}},
  \bibinfo {author} {\bibfnamefont {B.}~\bibnamefont {Malaescu}}, \ and\
  \bibinfo {author} {\bibfnamefont {Z.}~\bibnamefont {Zhang}},\ }\href
  {\doibase 10.1140/epjc/s10052-012-1874-8, 10.1140/epjc/s10052-010-1515-z}
  {\bibfield  {journal} {\bibinfo  {journal} {Eur. Phys. J.}\ }\textbf
  {\bibinfo {volume} {C71}},\ \bibinfo {pages} {1515} (\bibinfo {year}
  {2011})},\ \bibinfo {note} {[Erratum: Eur. Phys. J.C72,1874(2012)]},\ \Eprint
  {http://arxiv.org/abs/1010.4180} {arXiv:1010.4180 [hep-ph]} \BibitemShut
  {NoStop}%
\bibitem [{\citenamefont {Hagiwara}\ \emph {et~al.}(2011)\citenamefont
  {Hagiwara}, \citenamefont {Liao}, \citenamefont {Martin}, \citenamefont
  {Nomura},\ and\ \citenamefont {Teubner}}]{Hagiwara:2011af}%
  \BibitemOpen
  \bibfield  {author} {\bibinfo {author} {\bibfnamefont {K.}~\bibnamefont
  {Hagiwara}}, \bibinfo {author} {\bibfnamefont {R.}~\bibnamefont {Liao}},
  \bibinfo {author} {\bibfnamefont {A.~D.}\ \bibnamefont {Martin}}, \bibinfo
  {author} {\bibfnamefont {D.}~\bibnamefont {Nomura}}, \ and\ \bibinfo {author}
  {\bibfnamefont {T.}~\bibnamefont {Teubner}},\ }\href {\doibase
  10.1088/0954-3899/38/8/085003} {\bibfield  {journal} {\bibinfo  {journal}
  {J.Phys.}\ }\textbf {\bibinfo {volume} {G38}},\ \bibinfo {pages} {085003}
  (\bibinfo {year} {2011})},\ \Eprint {http://arxiv.org/abs/1105.3149}
  {arXiv:1105.3149 [hep-ph]} \BibitemShut {NoStop}%
\bibitem [{\citenamefont {Bennett}\ \emph {et~al.}(2006)\citenamefont {Bennett}
  \emph {et~al.}}]{Bennett:2006fi}%
  \BibitemOpen
  \bibfield  {author} {\bibinfo {author} {\bibfnamefont {G.}~\bibnamefont
  {Bennett}} \emph {et~al.} (\bibinfo {collaboration} {Muon G-2}),\ }\href
  {\doibase 10.1103/PhysRevD.73.072003} {\bibfield  {journal} {\bibinfo
  {journal} {Phys.Rev.}\ }\textbf {\bibinfo {volume} {D73}},\ \bibinfo {pages}
  {072003} (\bibinfo {year} {2006})},\ \Eprint
  {http://arxiv.org/abs/hep-ex/0602035} {arXiv:hep-ex/0602035 [hep-ex]}
  \BibitemShut {NoStop}%
\bibitem [{\citenamefont {Beringer}\ \emph {et~al.}(2012)\citenamefont
  {Beringer} \emph {et~al.}}]{PDG2013}%
  \BibitemOpen
  \bibfield  {author} {\bibinfo {author} {\bibfnamefont {J.}~\bibnamefont
  {Beringer}} \emph {et~al.} (\bibinfo {collaboration} {Particle Data Group}),\
  }\href@noop {} {\bibfield  {journal} {\bibinfo  {journal} {Phys. Rev.}\
  }\textbf {\bibinfo {volume} {D86}},\ \bibinfo {pages} {010001} (\bibinfo
  {year} {2012})},\ \bibinfo {note} {including the 2013 update for the 2014
  edition at http://pdg.lbl.gov}\BibitemShut {NoStop}%
\bibitem [{\citenamefont {Carey}\ \emph {et~al.}(2009)\citenamefont {Carey},
  \citenamefont {Lynch}, \citenamefont {Miller}, \citenamefont {Roberts},
  \citenamefont {Morse} \emph {et~al.}}]{Carey:2009zzb}%
  \BibitemOpen
  \bibfield  {author} {\bibinfo {author} {\bibfnamefont {R.}~\bibnamefont
  {Carey}}, \bibinfo {author} {\bibfnamefont {K.}~\bibnamefont {Lynch}},
  \bibinfo {author} {\bibfnamefont {J.}~\bibnamefont {Miller}}, \bibinfo
  {author} {\bibfnamefont {B.}~\bibnamefont {Roberts}}, \bibinfo {author}
  {\bibfnamefont {W.}~\bibnamefont {Morse}},  \emph {et~al.},\ }\href@noop {}
  {\enquote {\bibinfo {title} {{The New (g-2) Experiment: A proposal to measure
  the muon anomalous magnetic moment to +-0.14 ppm precision}},}\ } (\bibinfo
  {year} {2009})\BibitemShut {NoStop}%
\bibitem [{\citenamefont {Aoki}\ \emph {et~al.}(2009)\citenamefont {Aoki} \emph
  {et~al.}}]{Aoki:2009xxx}%
  \BibitemOpen
  \bibfield  {author} {\bibinfo {author} {\bibfnamefont {M.}~\bibnamefont
  {Aoki}} \emph {et~al.},\ }\href@noop {} {\bibfield  {journal} {\bibinfo
  {journal} {KEK-J-PARC}\ }\textbf {\bibinfo {volume} {PAC2009}},\ \bibinfo
  {pages} {12} (\bibinfo {year} {2009})}\BibitemShut {NoStop}%
\bibitem [{\citenamefont {Grange}\ \emph {et~al.}(2015)\citenamefont {Grange}
  \emph {et~al.}}]{Grange:2015fou}%
  \BibitemOpen
  \bibfield  {author} {\bibinfo {author} {\bibfnamefont {J.}~\bibnamefont
  {Grange}} \emph {et~al.} (\bibinfo {collaboration} {Muon g-2}),\ }\href@noop
  {} {\  (\bibinfo {year} {2015})},\ \Eprint {http://arxiv.org/abs/1501.06858}
  {arXiv:1501.06858 [physics.ins-det]} \BibitemShut {NoStop}%
\bibitem [{\citenamefont {Kurz}\ \emph {et~al.}(2014)\citenamefont {Kurz},
  \citenamefont {Liu}, \citenamefont {Marquard},\ and\ \citenamefont
  {Steinhauser}}]{Kurz:2014wya}%
  \BibitemOpen
  \bibfield  {author} {\bibinfo {author} {\bibfnamefont {A.}~\bibnamefont
  {Kurz}}, \bibinfo {author} {\bibfnamefont {T.}~\bibnamefont {Liu}}, \bibinfo
  {author} {\bibfnamefont {P.}~\bibnamefont {Marquard}}, \ and\ \bibinfo
  {author} {\bibfnamefont {M.}~\bibnamefont {Steinhauser}},\ }\href {\doibase
  10.1016/j.physletb.2014.05.043} {\bibfield  {journal} {\bibinfo  {journal}
  {Phys. Lett.}\ }\textbf {\bibinfo {volume} {B734}},\ \bibinfo {pages} {144}
  (\bibinfo {year} {2014})},\ \Eprint {http://arxiv.org/abs/1403.6400}
  {arXiv:1403.6400 [hep-ph]} \BibitemShut {NoStop}%
\bibitem [{\citenamefont {Prades}\ \emph {et~al.}(2009)\citenamefont {Prades},
  \citenamefont {de~Rafael},\ and\ \citenamefont {Vainshtein}}]{Prades:2009tw}%
  \BibitemOpen
  \bibfield  {author} {\bibinfo {author} {\bibfnamefont {J.}~\bibnamefont
  {Prades}}, \bibinfo {author} {\bibfnamefont {E.}~\bibnamefont {de~Rafael}}, \
  and\ \bibinfo {author} {\bibfnamefont {A.}~\bibnamefont {Vainshtein}},\
  }\href {\doibase 10.1142/9789814271844_0009} {\bibfield  {journal} {\bibinfo
  {journal} {Adv. Ser. Direct. High Energy Phys.}\ }\textbf {\bibinfo {volume}
  {20}},\ \bibinfo {pages} {303} (\bibinfo {year} {2009})},\ \Eprint
  {http://arxiv.org/abs/0901.0306} {arXiv:0901.0306 [hep-ph]} \BibitemShut
  {NoStop}%
\bibitem [{\citenamefont {Blum}\ \emph {et~al.}(2015)\citenamefont {Blum},
  \citenamefont {Christ}, \citenamefont {Hayakawa}, \citenamefont {Izubuchi},
  \citenamefont {Jin},\ and\ \citenamefont {Lehner}}]{Blum:2015gfa}%
  \BibitemOpen
  \bibfield  {author} {\bibinfo {author} {\bibfnamefont {T.}~\bibnamefont
  {Blum}}, \bibinfo {author} {\bibfnamefont {N.}~\bibnamefont {Christ}},
  \bibinfo {author} {\bibfnamefont {M.}~\bibnamefont {Hayakawa}}, \bibinfo
  {author} {\bibfnamefont {T.}~\bibnamefont {Izubuchi}}, \bibinfo {author}
  {\bibfnamefont {L.}~\bibnamefont {Jin}}, \ and\ \bibinfo {author}
  {\bibfnamefont {C.}~\bibnamefont {Lehner}},\ }\href@noop {} {\  (\bibinfo
  {year} {2015})},\ \Eprint {http://arxiv.org/abs/1510.07100} {arXiv:1510.07100
  [hep-lat]} \BibitemShut {NoStop}%
\bibitem [{\citenamefont {Green}\ \emph {et~al.}(2015)\citenamefont {Green},
  \citenamefont {Gryniuk}, \citenamefont {von Hippel}, \citenamefont {Meyer},\
  and\ \citenamefont {Pascalutsa}}]{Green:2015sra}%
  \BibitemOpen
  \bibfield  {author} {\bibinfo {author} {\bibfnamefont {J.}~\bibnamefont
  {Green}}, \bibinfo {author} {\bibfnamefont {O.}~\bibnamefont {Gryniuk}},
  \bibinfo {author} {\bibfnamefont {G.}~\bibnamefont {von Hippel}}, \bibinfo
  {author} {\bibfnamefont {H.~B.}\ \bibnamefont {Meyer}}, \ and\ \bibinfo
  {author} {\bibfnamefont {V.}~\bibnamefont {Pascalutsa}},\ }\href {\doibase
  10.1103/PhysRevLett.115.222003} {\bibfield  {journal} {\bibinfo  {journal}
  {Phys. Rev. Lett.}\ }\textbf {\bibinfo {volume} {115}},\ \bibinfo {pages}
  {222003} (\bibinfo {year} {2015})},\ \Eprint
  {http://arxiv.org/abs/1507.01577} {arXiv:1507.01577 [hep-lat]} \BibitemShut
  {NoStop}%
\bibitem [{\citenamefont {Blum}\ \emph
  {et~al.}(2014{\natexlab{a}})\citenamefont {Blum}, \citenamefont {Chowdhury},
  \citenamefont {Hayakawa},\ and\ \citenamefont {Izubuchi}}]{Blum:2014oka}%
  \BibitemOpen
  \bibfield  {author} {\bibinfo {author} {\bibfnamefont {T.}~\bibnamefont
  {Blum}}, \bibinfo {author} {\bibfnamefont {S.}~\bibnamefont {Chowdhury}},
  \bibinfo {author} {\bibfnamefont {M.}~\bibnamefont {Hayakawa}}, \ and\
  \bibinfo {author} {\bibfnamefont {T.}~\bibnamefont {Izubuchi}},\ }\href@noop
  {} {\  (\bibinfo {year} {2014}{\natexlab{a}})},\ \Eprint
  {http://arxiv.org/abs/1407.2923} {arXiv:1407.2923 [hep-lat]} \BibitemShut
  {NoStop}%
\bibitem [{\citenamefont {Gulpers}\ \emph {et~al.}(2014)\citenamefont
  {Gulpers}, \citenamefont {Francis}, \citenamefont {Jager}, \citenamefont
  {Meyer}, \citenamefont {von Hippel},\ and\ \citenamefont
  {Wittig}}]{Francis:2014hoa}%
  \BibitemOpen
  \bibfield  {author} {\bibinfo {author} {\bibfnamefont {V.}~\bibnamefont
  {Gulpers}}, \bibinfo {author} {\bibfnamefont {A.}~\bibnamefont {Francis}},
  \bibinfo {author} {\bibfnamefont {B.}~\bibnamefont {Jager}}, \bibinfo
  {author} {\bibfnamefont {H.}~\bibnamefont {Meyer}}, \bibinfo {author}
  {\bibfnamefont {G.}~\bibnamefont {von Hippel}}, \ and\ \bibinfo {author}
  {\bibfnamefont {H.}~\bibnamefont {Wittig}},\ }\bibfield  {booktitle} {\emph
  {\bibinfo {booktitle} {{Proceedings, 32nd International Symposium on Lattice
  Field Theory (Lattice 2014)}}},\ }\href@noop {} {\bibfield  {journal}
  {\bibinfo  {journal} {PoS}\ }\textbf {\bibinfo {volume} {LATTICE2014}},\
  \bibinfo {pages} {128} (\bibinfo {year} {2014})},\ \Eprint
  {http://arxiv.org/abs/1411.7592} {arXiv:1411.7592 [hep-lat]} \BibitemShut
  {NoStop}%
\bibitem [{\citenamefont {Bali}\ and\ \citenamefont
  {Endrodi}(2015)}]{Bali:2015msa}%
  \BibitemOpen
  \bibfield  {author} {\bibinfo {author} {\bibfnamefont {G.}~\bibnamefont
  {Bali}}\ and\ \bibinfo {author} {\bibfnamefont {G.}~\bibnamefont {Endrodi}},\
  }\href {\doibase 10.1103/PhysRevD.92.054506} {\bibfield  {journal} {\bibinfo
  {journal} {Phys. Rev.}\ }\textbf {\bibinfo {volume} {D92}},\ \bibinfo {pages}
  {054506} (\bibinfo {year} {2015})},\ \Eprint
  {http://arxiv.org/abs/1506.08638} {arXiv:1506.08638 [hep-lat]} \BibitemShut
  {NoStop}%
\bibitem [{\citenamefont {Burger}\ \emph {et~al.}(2015)\citenamefont {Burger},
  \citenamefont {Hotzel}, \citenamefont {Jansen},\ and\ \citenamefont
  {Petschlies}}]{Burger:2015oya}%
  \BibitemOpen
  \bibfield  {author} {\bibinfo {author} {\bibfnamefont {F.}~\bibnamefont
  {Burger}}, \bibinfo {author} {\bibfnamefont {G.}~\bibnamefont {Hotzel}},
  \bibinfo {author} {\bibfnamefont {K.}~\bibnamefont {Jansen}}, \ and\ \bibinfo
  {author} {\bibfnamefont {M.}~\bibnamefont {Petschlies}},\ }\href@noop {} {\
  (\bibinfo {year} {2015})},\ \Eprint {http://arxiv.org/abs/1501.05110}
  {arXiv:1501.05110 [hep-lat]} \BibitemShut {NoStop}%
\bibitem [{\citenamefont {Chakraborty}\ \emph {et~al.}(2015)\citenamefont
  {Chakraborty}, \citenamefont {Davies}, \citenamefont {Koponen}, \citenamefont
  {Lepage}, \citenamefont {Peardon},\ and\ \citenamefont
  {Ryan}}]{Chakraborty:2015ugp}%
  \BibitemOpen
  \bibfield  {author} {\bibinfo {author} {\bibfnamefont {B.}~\bibnamefont
  {Chakraborty}}, \bibinfo {author} {\bibfnamefont {C.~T.~H.}\ \bibnamefont
  {Davies}}, \bibinfo {author} {\bibfnamefont {J.}~\bibnamefont {Koponen}},
  \bibinfo {author} {\bibfnamefont {G.~P.}\ \bibnamefont {Lepage}}, \bibinfo
  {author} {\bibfnamefont {M.~J.}\ \bibnamefont {Peardon}}, \ and\ \bibinfo
  {author} {\bibfnamefont {S.~M.}\ \bibnamefont {Ryan}},\ }\href@noop {} {\
  (\bibinfo {year} {2015})},\ \Eprint {http://arxiv.org/abs/1512.03270}
  {arXiv:1512.03270 [hep-lat]} \BibitemShut {NoStop}%
\bibitem [{\citenamefont {Toth}(2015)}]{BMWDiscoLattice2015}%
  \BibitemOpen
  \bibfield  {author} {\bibinfo {author} {\bibfnamefont {B.}~\bibnamefont
  {Toth}},\ }\href@noop {} {\enquote {\bibinfo {title} {{Disconnected
  contribution to hadron correlation functions}},}\ } (\bibinfo {year}
  {2015}),\ \bibinfo {note} {the 33rd International Symposium on Lattice Field
  Theory.}\BibitemShut {Stop}%
\bibitem [{\citenamefont {Blum}(2003)}]{Blum:2002ii}%
  \BibitemOpen
  \bibfield  {author} {\bibinfo {author} {\bibfnamefont {T.}~\bibnamefont
  {Blum}},\ }\href {\doibase 10.1103/PhysRevLett.91.052001} {\bibfield
  {journal} {\bibinfo  {journal} {Phys.Rev.Lett.}\ }\textbf {\bibinfo {volume}
  {91}},\ \bibinfo {pages} {052001} (\bibinfo {year} {2003})},\ \Eprint
  {http://arxiv.org/abs/hep-lat/0212018} {arXiv:hep-lat/0212018 [hep-lat]}
  \BibitemShut {NoStop}%
\bibitem [{\citenamefont {Bernecker}\ and\ \citenamefont
  {Meyer}(2011)}]{Bernecker:2011gh}%
  \BibitemOpen
  \bibfield  {author} {\bibinfo {author} {\bibfnamefont {D.}~\bibnamefont
  {Bernecker}}\ and\ \bibinfo {author} {\bibfnamefont {H.~B.}\ \bibnamefont
  {Meyer}},\ }\href {\doibase 10.1140/epja/i2011-11148-6} {\bibfield  {journal}
  {\bibinfo  {journal} {Eur. Phys. J.}\ }\textbf {\bibinfo {volume} {A47}},\
  \bibinfo {pages} {148} (\bibinfo {year} {2011})},\ \Eprint
  {http://arxiv.org/abs/1107.4388} {arXiv:1107.4388 [hep-lat]} \BibitemShut
  {NoStop}%
\bibitem [{\citenamefont {Feng}\ \emph {et~al.}(2013)\citenamefont {Feng},
  \citenamefont {Hashimoto}, \citenamefont {Hotzel}, \citenamefont {Jansen},
  \citenamefont {Petschlies} \emph {et~al.}}]{Feng:2013xsa}%
  \BibitemOpen
  \bibfield  {author} {\bibinfo {author} {\bibfnamefont {X.}~\bibnamefont
  {Feng}}, \bibinfo {author} {\bibfnamefont {S.}~\bibnamefont {Hashimoto}},
  \bibinfo {author} {\bibfnamefont {G.}~\bibnamefont {Hotzel}}, \bibinfo
  {author} {\bibfnamefont {K.}~\bibnamefont {Jansen}}, \bibinfo {author}
  {\bibfnamefont {M.}~\bibnamefont {Petschlies}},  \emph {et~al.},\ }\href
  {\doibase 10.1103/PhysRevD.88.034505} {\bibfield  {journal} {\bibinfo
  {journal} {Phys.Rev.}\ }\textbf {\bibinfo {volume} {D88}},\ \bibinfo {pages}
  {034505} (\bibinfo {year} {2013})},\ \Eprint {http://arxiv.org/abs/1305.5878}
  {arXiv:1305.5878 [hep-lat]} \BibitemShut {NoStop}%
\bibitem [{Note1()}]{Note1}%
  \BibitemOpen
  \bibinfo {note} {The appropriate normalization factors of $Z_V^2$ are of
  course included in our computation.}\BibitemShut {Stop}%
\bibitem [{\citenamefont {Foley}\ \emph {et~al.}(2005)\citenamefont {Foley},
  \citenamefont {Jimmy~Juge}, \citenamefont {O'Cais}, \citenamefont {Peardon},
  \citenamefont {Ryan},\ and\ \citenamefont {Skullerud}}]{Foley:2005ac}%
  \BibitemOpen
  \bibfield  {author} {\bibinfo {author} {\bibfnamefont {J.}~\bibnamefont
  {Foley}}, \bibinfo {author} {\bibfnamefont {K.}~\bibnamefont {Jimmy~Juge}},
  \bibinfo {author} {\bibfnamefont {A.}~\bibnamefont {O'Cais}}, \bibinfo
  {author} {\bibfnamefont {M.}~\bibnamefont {Peardon}}, \bibinfo {author}
  {\bibfnamefont {S.~M.}\ \bibnamefont {Ryan}}, \ and\ \bibinfo {author}
  {\bibfnamefont {J.-I.}\ \bibnamefont {Skullerud}},\ }\href {\doibase
  10.1016/j.cpc.2005.06.008} {\bibfield  {journal} {\bibinfo  {journal}
  {Comput. Phys. Commun.}\ }\textbf {\bibinfo {volume} {172}},\ \bibinfo
  {pages} {145} (\bibinfo {year} {2005})},\ \Eprint
  {http://arxiv.org/abs/hep-lat/0505023} {arXiv:hep-lat/0505023 [hep-lat]}
  \BibitemShut {NoStop}%
\bibitem [{\citenamefont {Li}\ \emph {et~al.}(2010)\citenamefont {Li} \emph
  {et~al.}}]{Li:2010pw}%
  \BibitemOpen
  \bibfield  {author} {\bibinfo {author} {\bibfnamefont {A.}~\bibnamefont {Li}}
  \emph {et~al.} (\bibinfo {collaboration} {xQCD}),\ }\href {\doibase
  10.1103/PhysRevD.82.114501} {\bibfield  {journal} {\bibinfo  {journal} {Phys.
  Rev.}\ }\textbf {\bibinfo {volume} {D82}},\ \bibinfo {pages} {114501}
  (\bibinfo {year} {2010})},\ \Eprint {http://arxiv.org/abs/1005.5424}
  {arXiv:1005.5424 [hep-lat]} \BibitemShut {NoStop}%
\bibitem [{\citenamefont {Brower}\ \emph {et~al.}(2005)\citenamefont {Brower},
  \citenamefont {Neff},\ and\ \citenamefont {Orginos}}]{Brower:2004xi}%
  \BibitemOpen
  \bibfield  {author} {\bibinfo {author} {\bibfnamefont {R.~C.}\ \bibnamefont
  {Brower}}, \bibinfo {author} {\bibfnamefont {H.}~\bibnamefont {Neff}}, \ and\
  \bibinfo {author} {\bibfnamefont {K.}~\bibnamefont {Orginos}},\ }\bibfield
  {booktitle} {\emph {\bibinfo {booktitle} {{Lattice field theory. Proceedings,
  22nd International Symposium, Lattice 2004, Batavia, USA, June 21-26,
  2004}}},\ }\href {\doibase 10.1016/j.nuclphysbps.2004.11.180} {\bibfield
  {journal} {\bibinfo  {journal} {Nucl. Phys. Proc. Suppl.}\ }\textbf {\bibinfo
  {volume} {140}},\ \bibinfo {pages} {686} (\bibinfo {year} {2005})},\ \bibinfo
  {note} {[,686(2004)]},\ \Eprint {http://arxiv.org/abs/hep-lat/0409118}
  {arXiv:hep-lat/0409118 [hep-lat]} \BibitemShut {NoStop}%
\bibitem [{\citenamefont {Blum}\ \emph
  {et~al.}(2014{\natexlab{b}})\citenamefont {Blum} \emph
  {et~al.}}]{Blum:2014tka}%
  \BibitemOpen
  \bibfield  {author} {\bibinfo {author} {\bibfnamefont {T.}~\bibnamefont
  {Blum}} \emph {et~al.} (\bibinfo {collaboration} {RBC, UKQCD}),\ }\href@noop
  {} {\  (\bibinfo {year} {2014}{\natexlab{b}})},\ \Eprint
  {http://arxiv.org/abs/1411.7017} {arXiv:1411.7017 [hep-lat]} \BibitemShut
  {NoStop}%
\bibitem [{\citenamefont {Blum}\ \emph {et~al.}(2012)\citenamefont {Blum},
  \citenamefont {Izubuchi},\ and\ \citenamefont {Shintani}}]{Blum:2012uh}%
  \BibitemOpen
  \bibfield  {author} {\bibinfo {author} {\bibfnamefont {T.}~\bibnamefont
  {Blum}}, \bibinfo {author} {\bibfnamefont {T.}~\bibnamefont {Izubuchi}}, \
  and\ \bibinfo {author} {\bibfnamefont {E.}~\bibnamefont {Shintani}},\
  }\href@noop {} {\  (\bibinfo {year} {2012})},\ \Eprint
  {http://arxiv.org/abs/1208.4349} {arXiv:1208.4349 [hep-lat]} \BibitemShut
  {NoStop}%
\bibitem [{\citenamefont {Shintani}\ \emph {et~al.}(2014)\citenamefont
  {Shintani}, \citenamefont {Arthur}, \citenamefont {Blum}, \citenamefont
  {Izubuchi}, \citenamefont {Jung},\ and\ \citenamefont
  {Lehner}}]{Shintani:2014vja}%
  \BibitemOpen
  \bibfield  {author} {\bibinfo {author} {\bibfnamefont {E.}~\bibnamefont
  {Shintani}}, \bibinfo {author} {\bibfnamefont {R.}~\bibnamefont {Arthur}},
  \bibinfo {author} {\bibfnamefont {T.}~\bibnamefont {Blum}}, \bibinfo {author}
  {\bibfnamefont {T.}~\bibnamefont {Izubuchi}}, \bibinfo {author}
  {\bibfnamefont {C.}~\bibnamefont {Jung}}, \ and\ \bibinfo {author}
  {\bibfnamefont {C.}~\bibnamefont {Lehner}},\ }\href@noop {} {\  (\bibinfo
  {year} {2014})},\ \Eprint {http://arxiv.org/abs/1402.0244} {arXiv:1402.0244
  [hep-lat]} \BibitemShut {NoStop}%
\bibitem [{\citenamefont {Jung}(2015)}]{ChulwooLattice2015}%
  \BibitemOpen
  \bibfield  {author} {\bibinfo {author} {\bibfnamefont {C.}~\bibnamefont
  {Jung}},\ }\href@noop {} {\enquote {\bibinfo {title} {{zMobius and other
  recent developments on Domain Wall Fermions}},}\ } (\bibinfo {year} {2015}),\
  \bibinfo {note} {the 33rd International Symposium on Lattice Field
  Theory.}\BibitemShut {Stop}%
\bibitem [{\citenamefont {Della~Morte}\ and\ \citenamefont
  {J{\"u}ttner}(2010)}]{DellaMorte:2010aq}%
  \BibitemOpen
  \bibfield  {author} {\bibinfo {author} {\bibfnamefont {M.}~\bibnamefont
  {Della~Morte}}\ and\ \bibinfo {author} {\bibfnamefont {A.}~\bibnamefont
  {J{\"u}ttner}},\ }\href {\doibase 10.1007/JHEP11(2010)154} {\bibfield
  {journal} {\bibinfo  {journal} {JHEP}\ }\textbf {\bibinfo {volume} {11}},\
  \bibinfo {pages} {154} (\bibinfo {year} {2010})},\ \Eprint
  {http://arxiv.org/abs/1009.3783} {arXiv:1009.3783 [hep-lat]} \BibitemShut
  {NoStop}%
\bibitem [{Note2()}]{Note2}%
  \BibitemOpen
  \bibinfo {note} {Alternatively taking $T=21$ instead of $T=20$ and repeating
  our procedure to estimate systematic uncertainties, we find $a_\mu ^{\protect
  \rm HVP~(LO)~DISC} = -8.3(4.0)(1.8)\times 10^{-10}$, where the first error is
  statistical and the second systematic. This value is consistent with our
  preferred value, however, has a different balance of statistical and
  systematic errors.}\BibitemShut {Stop}%
\bibitem [{\citenamefont {Spraggs}(2015)}]{MattLattice2015}%
  \BibitemOpen
  \bibfield  {author} {\bibinfo {author} {\bibfnamefont {M.}~\bibnamefont
  {Spraggs}},\ }\href@noop {} {\enquote {\bibinfo {title} {{The strange
  contribution to $a^{\rm HVP,LO}_\mu$ with physical quark masses using Mobius
  domain wall fermions}},}\ } (\bibinfo {year} {2015}),\ \bibinfo {note} {the
  33rd International Symposium on Lattice Field Theory.}\BibitemShut {Stop}%
\bibitem [{\citenamefont {Aubin}\ \emph {et~al.}(2015)\citenamefont {Aubin},
  \citenamefont {Blum}, \citenamefont {Chau}, \citenamefont {Golterman},
  \citenamefont {Peris},\ and\ \citenamefont {Tu}}]{Aubin:2015rzx}%
  \BibitemOpen
  \bibfield  {author} {\bibinfo {author} {\bibfnamefont {C.}~\bibnamefont
  {Aubin}}, \bibinfo {author} {\bibfnamefont {T.}~\bibnamefont {Blum}},
  \bibinfo {author} {\bibfnamefont {P.}~\bibnamefont {Chau}}, \bibinfo {author}
  {\bibfnamefont {M.}~\bibnamefont {Golterman}}, \bibinfo {author}
  {\bibfnamefont {S.}~\bibnamefont {Peris}}, \ and\ \bibinfo {author}
  {\bibfnamefont {C.}~\bibnamefont {Tu}},\ }\href@noop {} {\  (\bibinfo {year}
  {2015})},\ \Eprint {http://arxiv.org/abs/1512.07555} {arXiv:1512.07555
  [hep-lat]} \BibitemShut {NoStop}%
\end{thebibliography}%

\clearpage

\setcounter{page}{1}
\renewcommand{\thepage}{Supplementary Information -- S\arabic{page}}
\setcounter{table}{0}
\renewcommand{\thetable}{S\,\Roman{table}}
\setcounter{equation}{0}
\renewcommand{\theequation}{S\,\arabic{equation}}

\begin{figure}[htbp]
\begin{center}
\includegraphics[scale=\figscale,page=1]{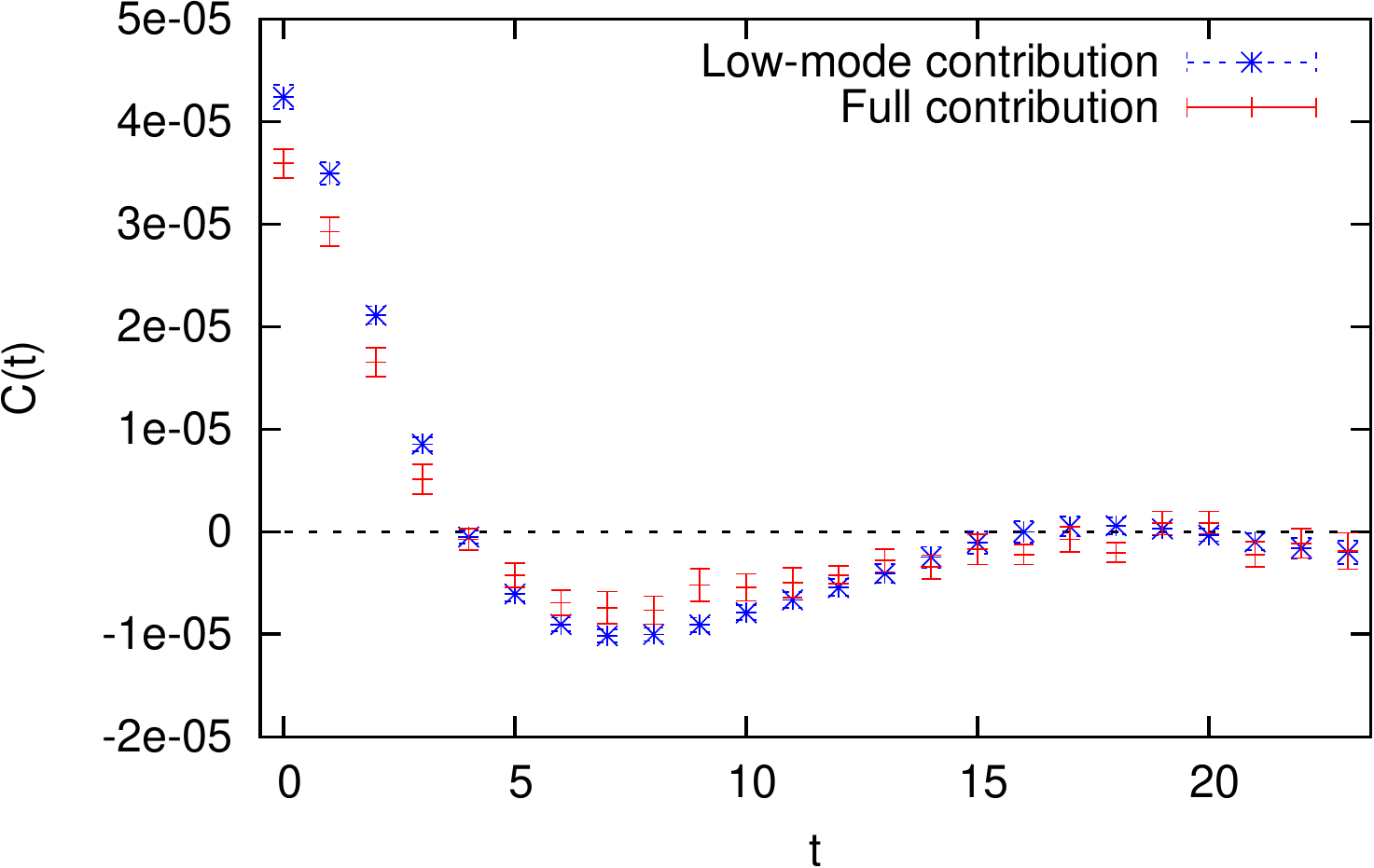}
\end{center} 
\caption{The low-mode contribution to $C(t)$ is contrasted with the
  full result.  As mentioned in the main text, the signal is mostly
  saturated by the low modes.}
\label{fig:lowm}
\end{figure}

\begin{figure}[htbp]
\begin{center}
\includegraphics[scale=\figscale,page=2]{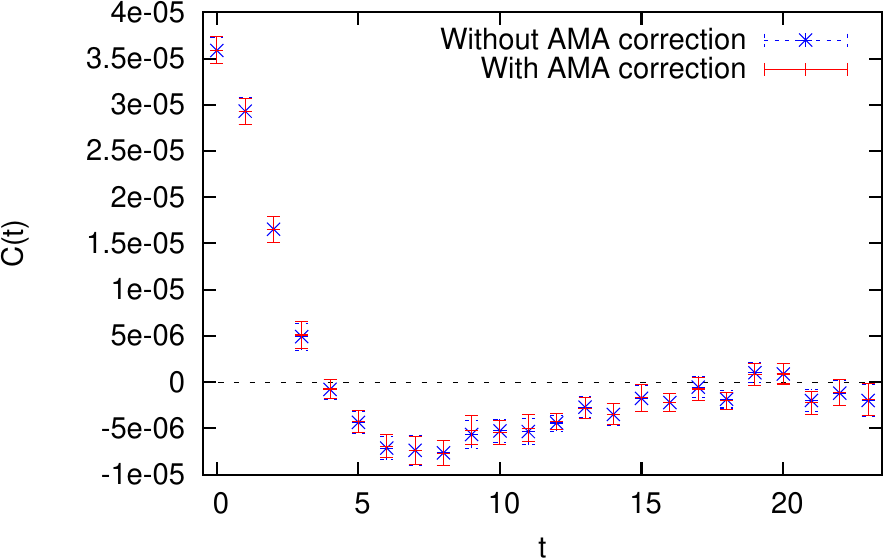}
\end{center} 
\caption{The correlator $C(t)$ is plotted with and without the AMA
  correction term.  The noise of the AMA estimator is dominated by the
  sloppy solves.}
\label{fig:ama}
\end{figure}

\begin{figure}[htbp]
\begin{center}
\includegraphics[scale=\figscale,page=4]{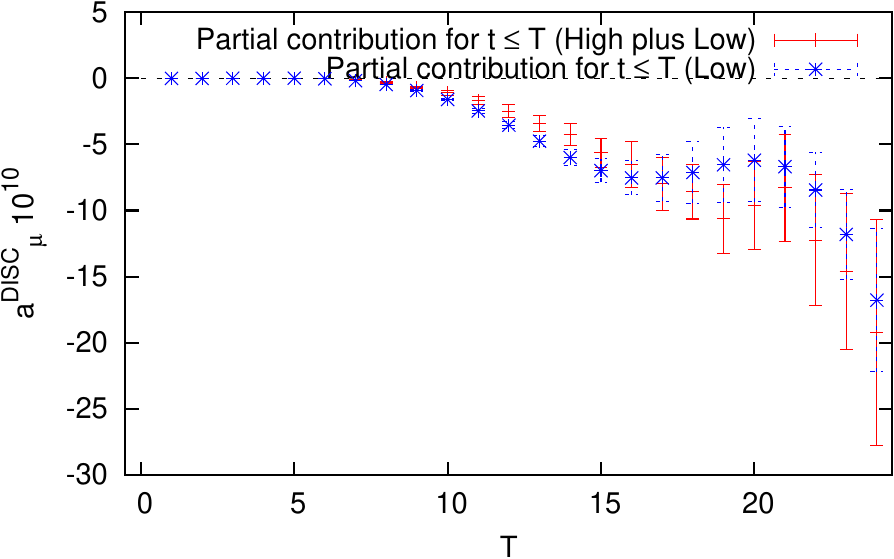}
\end{center} 
\caption{The partial sum $L_T$ is plotted with and without the high-mode
  contribution.  The similar uncertainty at around $T\approx 20$
  suggests that the additional high-mode noise is well controlled and
  that the noise is dominated by gauge-field fluctuations.}
\label{fig:znoisecan}
\end{figure}

\begin{figure}[htbp]
\begin{center}
\includegraphics[scale=\figscale,page=10]{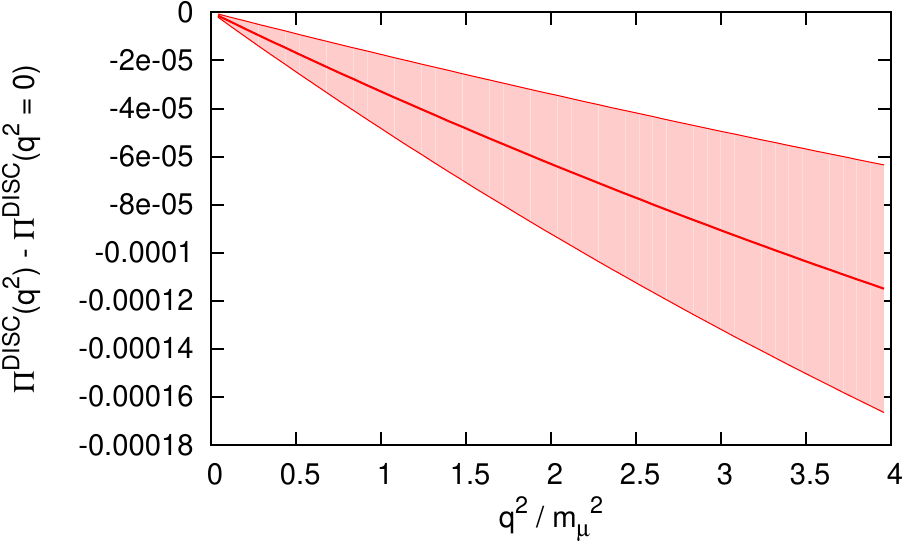}
\end{center} 
\caption{The subtracted scalar vacuum polarization function $\Pi^{\rm DISC}(q^2)$
  for the disconnected contribution from our physical-pion-mass
  lattice data is plotted in the interesting region of $q^2\in[0,4
  m_\mu^2]$.  The shaded band corresponds to the statistical
  uncertainty of our result.}
\label{fig:pihat}
\end{figure}

\begin{figure}[h!]
\begin{center}
\includegraphics[scale=\figscale,page=11]{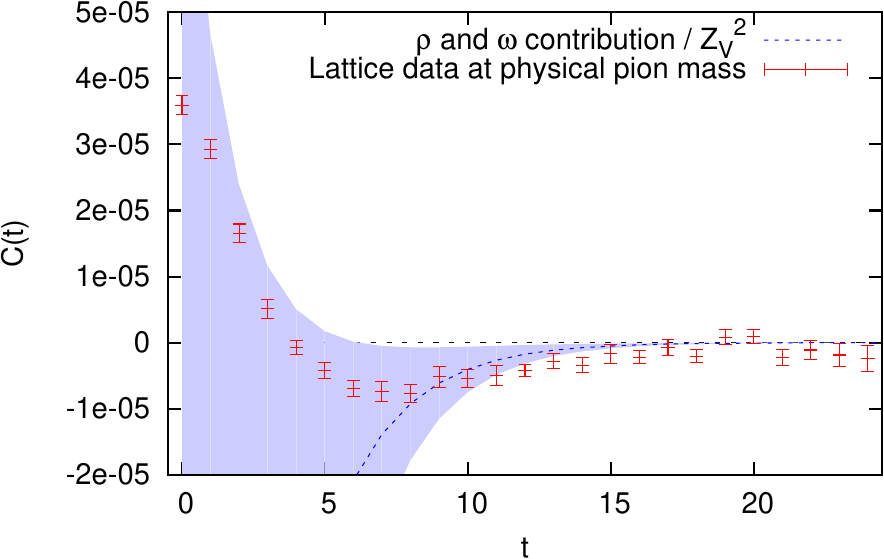}
\end{center} 
\caption{The long-distance model based on $\rho$ and $\omega$
  resonances used quantitatively in Ref.~\cite{Chakraborty:2015ugp} is
  shown on top of our lattice data at physical pion mass.  The shaded
  region corresponds to the parametric uncertainty of the model
  prediction with central value shown as dashed line.}
\label{fig:rhoomega}
\end{figure}

\begin{figure}[h!]
\begin{center}
\includegraphics[scale=\figscale,page=1]{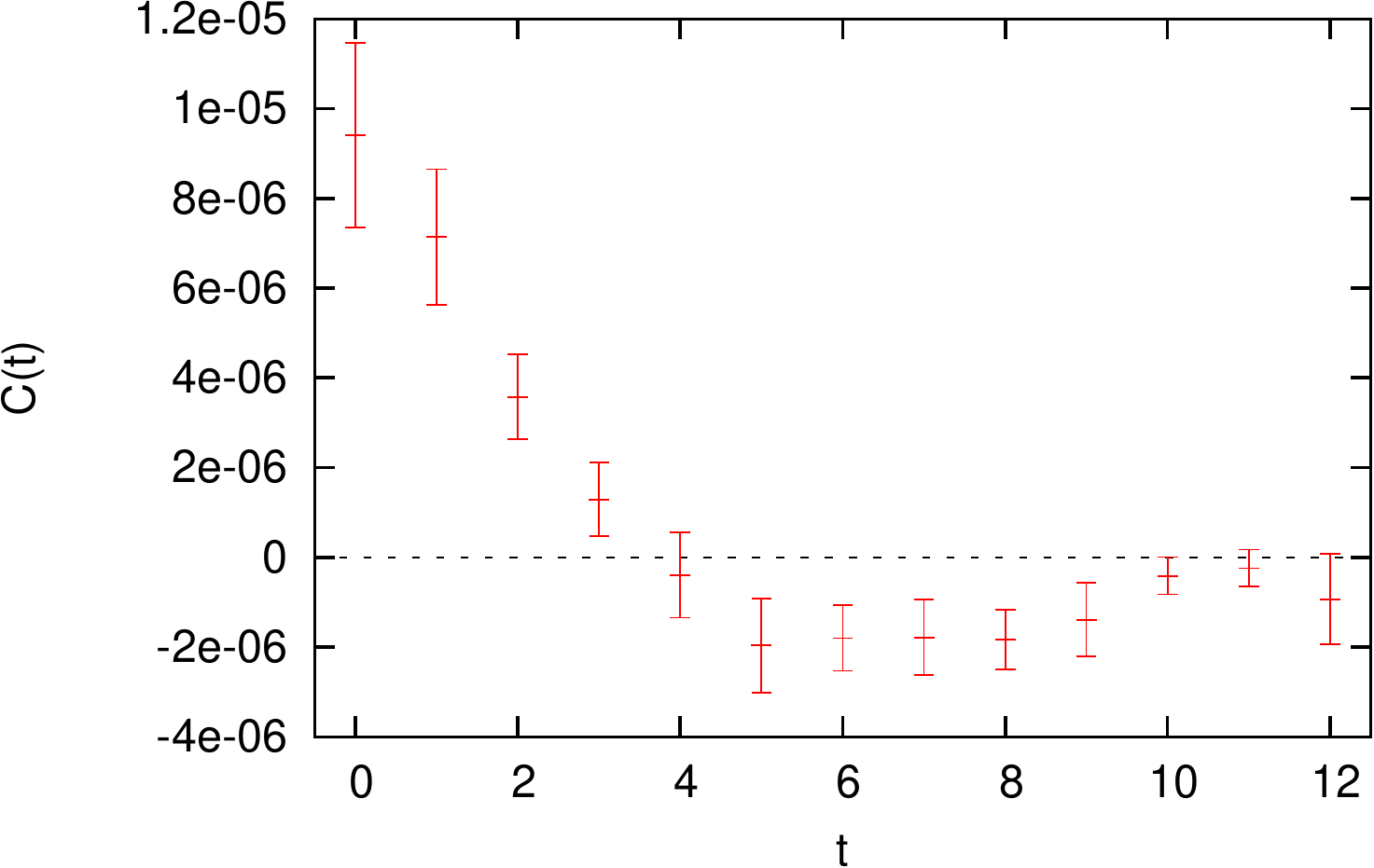}
\end{center} 
\caption{The correlator for heavy pion mass $m_\pi=422$ MeV.}
\label{fig:heavy}
\end{figure}

\begin{figure}[htbp]
\begin{center}
\includegraphics[scale=\figscale,page=2]{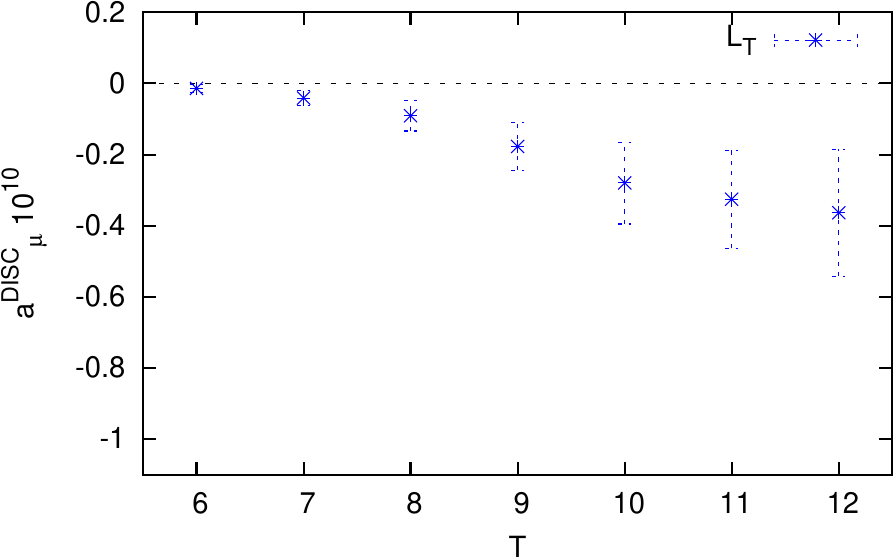}
\end{center} 
\caption{For heavy pion mass $m_\pi=422$ MeV, we find a very small
  result for the HVP disconnected contribution, $a_\mu \approx
  -0.4(2) \times 10^{-10}$, consistent with the results of
  Ref.~\cite{Chakraborty:2015ugp} at similar pion mass.}
\label{fig:heavyb}
\end{figure}

In the following we add additional plots supplementing relevant
technical details regarding our results.
Figs.~\ref{fig:lowm}--\ref{fig:rhoomega} are for physical pion mass.
Figs.~\ref{fig:heavy} and \ref{fig:heavyb} are for heavy pion mass
$m_\pi=422$ MeV.  The results for heavy pion mass are obtained from
only 6 configurations of the RBC and UKQCD $24^3 \times 64$ lattice.
The AMA setup uses 45 sloppy solves and 4 exact solves per
configuration \cite{Blum:2012uh,Shintani:2014vja}.

\clearpage
~

\clearpage

\end{document}